\documentclass[twocolumn]{aastex63}

\usepackage[T1]{fontenc}
\usepackage{ae,aecompl}
\usepackage{graphicx}	
\usepackage{amsmath}	
\usepackage{amssymb}

\received{2020 March 2}
\revised{2020 August 25}
\accepted{2020 September 26}
\submitjournal{ApJ}

\shorttitle{Modelling scintillation arcs of PSR~J0437$-$4715}
\shortauthors{Reardon et al.}

\begin{document}

\title{Precision orbital dynamics from interstellar scintillation arcs for PSR~J0437$-$4715}

\correspondingauthor{Daniel J. Reardon}
\email{dreardon@swin.edu.au}
\author{Daniel J. Reardon}
\affiliation{Centre for Astrophysics and Supercomputing, Swinburne University of Technology, P.O. Box 218, Hawthorn, Victoria 3122, Australia}
\affiliation{Australian Research Council Centre of Excellence for Gravitational Wave Discovery (OzGrav)}
\author{William A. Coles}
\affiliation{Electrical and Computer Engineering, University of California at San Diego, La Jolla, California, U.S.A.}

\author{Matthew Bailes}
\affiliation{Centre for Astrophysics and Supercomputing, Swinburne University of Technology, P.O. Box 218, Hawthorn, Victoria 3122, Australia}
\affiliation{Australian Research Council Centre of Excellence for Gravitational Wave Discovery (OzGrav)}
\author{N. D. Ramesh Bhat}
\affiliation{International Centre for Radio Astronomy Research, Curtin University, Bentley, Western Australia 6102, Australia}
\author{Shi Dai}
\affiliation{Australia Telescope National Facility, CSIRO Astronomy \& Space Science, P.O. Box 76, Epping, NSW 1710, Australia}
%
%
\author{George B. Hobbs}
\affiliation{Australia Telescope National Facility, CSIRO Astronomy \& Space Science, P.O. Box 76, Epping, NSW 1710, Australia}
\affiliation{Australian Research Council Centre of Excellence for Gravitational Wave Discovery (OzGrav)}
\author{Matthew Kerr}
\affiliation{Space Science Division, Naval Research Laboratory, Washington, DC 20375-5352, USA}

%
%
\author{Richard N. Manchester}
\affiliation{Australia Telescope National Facility, CSIRO Astronomy \& Space Science, P.O. Box 76, Epping, NSW 1710, Australia}
\author{Stefan Os{\l}owski}
\affiliation{Centre for Astrophysics and Supercomputing, Swinburne University of Technology, P.O. Box 218, Hawthorn, Victoria 3122, Australia}
\author{Aditya Parthasarathy}
\affiliation{Centre for Astrophysics and Supercomputing, Swinburne University of Technology, P.O. Box 218, Hawthorn, Victoria 3122, Australia}
\affiliation{Australian Research Council Centre of Excellence for Gravitational Wave Discovery (OzGrav)}

\author{Christopher J. Russell}
\affiliation{CSIRO Scientific Computing Services, Australian Technology Park, Locked Bag 9013, Alexandria, NSW 1435, Australia}
\author{Ryan M. Shannon}
\affiliation{Centre for Astrophysics and Supercomputing, Swinburne University of Technology, P.O. Box 218, Hawthorn, Victoria 3122, Australia}
\affiliation{Australian Research Council Centre of Excellence for Gravitational Wave Discovery (OzGrav)}

\author{Ren\'ee Spiewak}
\affiliation{Centre for Astrophysics and Supercomputing, Swinburne University of Technology, P.O. Box 218, Hawthorn, Victoria 3122, Australia}
\affiliation{Australian Research Council Centre of Excellence for Gravitational Wave Discovery (OzGrav)}
\author{Lawrence Toomey}
\affiliation{Australia Telescope National Facility, CSIRO Astronomy \& Space Science, P.O. Box 76, Epping, NSW 1710, Australia}
\author{Artem V. Tuntsov}
\affiliation{Manly Astrophysics, 15/41-42 East Esplanade, Manly, NSW 2095, Australia}
\author{Willem van Straten}
\affiliation{Institute for Radio Astronomy \& Space Research,
Auckland University of Technology, Private Bag 92006, Auckland 1142, New Zealand}
\author{Mark A. Walker}
\affiliation{Manly Astrophysics, 15/41-42 East Esplanade, Manly, NSW 2095, Australia}
\author{Jingbo Wang}
\affiliation{Xinjiang Astronomical Observatory, Chinese Academy of Sciences,
150 Science 1-Street, Urumqi, Xinjiang 830011, China}

\author{Lei Zhang}
\affiliation{National Astronomical Observatories, Chinese Academy of Sciences, A20 Datun Road, Chaoyang District, Beijing 100101, China}
\affiliation{University of Chinese Academy of Sciences, Beijing 100049, China}
\affiliation{Australia Telescope National Facility, CSIRO Astronomy \& Space Science, P.O. Box 76, Epping, NSW 1710, Australia}


\author{Xing-Jiang Zhu}
\affiliation{School of Physics and Astronomy, Monash University, Victoria 3800, Australia}
\affiliation{Australian Research Council Centre of Excellence for Gravitational Wave Discovery (OzGrav)}

\begin{abstract}
Intensity scintillations of radio pulsars are known to originate from interference between waves scattered by the electron density irregularities of interstellar plasma, often leading to parabolic arcs in the two-dimensional power spectrum of the recorded dynamic spectrum. The degree of arc curvature depends on the distance to the scattering plasma and its transverse velocity with respect to the line-of-sight. We report the observation of annual and orbital variations in the curvature of scintillation arcs over a period of $16$\,years for the bright millisecond pulsar, PSR~J0437$-$4715. These variations are the signature of the relative transverse motions of the Earth, pulsar, and scattering medium, which we model to obtain precise measurements of parameters of the pulsar's binary orbit and the scattering medium itself. We observe two clear scintillation arcs in most of our $>$5000 observations and we show that they originate from scattering by thin screens located at distances $D_1 = 89.8 \pm 0.4$\,pc and $D_2 = 124 \pm 3$\,pc from Earth. The best-fit scattering model we derive for the brightest arc yields the pulsar's orbital inclination angle $i = 137.1 \pm 0.3^\circ$, and longitude of ascending node, $\Omega=206.3\pm0.4^\circ$. Using scintillation arcs for precise astrometry and orbital dynamics can be superior to modelling variations in the diffractive scintillation timescale, because the arc curvature is independent of variations in the level of turbulence of interstellar plasma. This technique can be used in combination with pulsar timing to determine the full three-dimensional orbital geometries of binary pulsars, and provides parameters essential for testing theories of gravity and constraining neutron star masses. 
\end{abstract}
\vspace{10mm}
\section{Introduction} \label{sec:intro}

PSR~J0437$-$4715 is the nearest and brightest millisecond radio pulsar. It is a key part of the Parkes Pulsar Timing Array \citep[PPTA;][]{Manchester+13} project that monitors the arrival times of a set of millisecond pulsars over many years for the primary goal of detecting nanohertz-frequency gravitational waves. The pulsar has been observed with high cadence over the past 14 years as part of this project, and for a decade prior in legacy projects, because it is one of the most precisely-timed pulsars.

The pulsar timing technique has provided insights into relativistic dynamics \citep[e.g.][]{Kramer+06} through the study of pulsar orbits, but it is most sensitive to radial changes, meaning that for most pulsars only the projected orbit can be determined. Solving an orbit in three-dimensions requires a measurement of the orbital inclination $i$, which is difficult to achieve through timing unless the orbit is observed nearly edge-on or has extremely high precision arrival times \citep{vanStraten+01} to allow measurement of the relativistic Shapiro delay \citep{Shapiro64}. It can also be obtained from a subtle kinematic effect that depends on the Earth's orbit and pulsar proper motion \citep{Kopeikin95, Kopeikin96, vanStraten+01}, but this is only possible for the most precisely-timed millisecond pulsars, and PSR~J0437$-$4715 is one example. This kinematic term also depends on the longitude of ascending node $\Omega$, and is often a source of contamination in relativistic parameters such as periastron advance and the rate of change of the semi-major axis. A complimentary method for measuring $i$ and $\Omega$ is the study of pulsar scintillation, which in contrast to timing is only sensitive to transverse motions \citep[e.g.][]{Lyne84, Ord+02a, Ransom+04}.

\begin{figure*}
\centering
\includegraphics[width=.5454\textwidth]{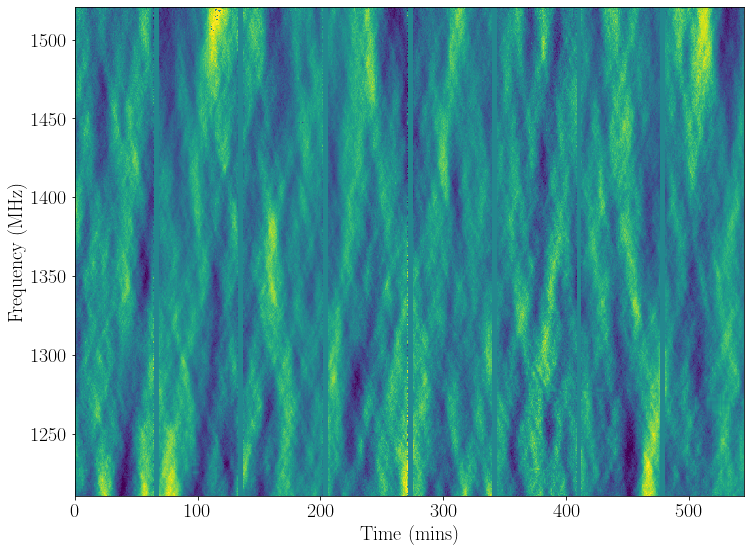}
\includegraphics[width=.4091\textwidth]{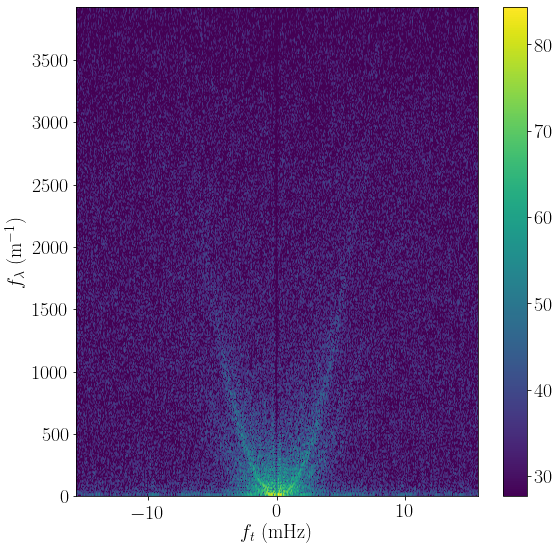}
\caption{Dynamic spectrum $S(t,\nu)$ (left) and corresponding secondary spectrum $P(f_t, f_\lambda)$ (right) for a long track observation of PSR~J0437$-$4715 from the Parkes 64-m radio telescope in the 20-cm observing band on MJD~55915. This dynamic spectrum shows eight consecutive observations, each separated by a $\sim$4\,minute gap during which a noise diode was observed for the purpose of polarisation calibration. The data in these gap has been replaced with the mean flux, and each frequency channel and sub-integration has been normalized to the same flux in order to show the fine-scale ``criss-cross" pattern that produces the scintillation arc phenomenon. The color scale in the dynamic spectrum is a linear scale for the flux in arbitrary units, while the scale in the secondary spectrum shows the power level in dB (with the low end of the scale being 3\,dB below the median and the high end being 3\,dB below the maximum).}
\label{fig:specs}
\end{figure*}

Interstellar scintillation originates from spatial fluctuations in the electron density of the ionized interstellar medium (IISM), which have a power-law distribution of sizes and densities originating from turbulence \citep{Rickett90}. These density fluctuations scatter incident wavefronts by diffraction and produce an interference pattern of intensity variations \citep{Rickett69} that varies with frequency, and with time because of the relative motions of source, scattering media, and observer. Scintillation is observed in radio observations of compact sources at centimetre to metre wavelengths and is captured in the dynamic spectrum (see left panel of Figure \ref{fig:specs}), which we describe in Section \ref{sec:observations}. 

The apparent quasi-periodic structures in the dynamic spectrum become more ordered in the secondary spectrum; the two-dimensional power spectrum of the dynamic spectrum (Figure \ref{fig:specs}, right panel). This may also be referred to as a delay-Doppler distribution because the Fourier conjugate variables on the axes correspond to the differential time delay $f_\nu$ and differential Doppler shift $f_t$ between pairs of interfering waves \citep{Cordes+06}. Parabolic arcs in a secondary spectrum were first identified by \citet{Stinebring+01}, and the origin of their shape was explained by \citet{Walker+04}.

The arcs shown here would be called ``forward'' arcs because the apex is at the origin and they can be described by 
$f_\nu = \eta_\nu f_t^2$.
Their parabolic form can be understood in terms of the interference of two plane waves scattered at angles $\theta_1$ and $\theta_2$. The Doppler shift,
$f_t\propto\theta_{1}-\theta_2$ and the delay, $f_\nu\propto \theta_{1}^2-\theta_2^2$. If one of $\theta_1$ or $\theta_2$ is near the origin a forward arc arises naturally. In strong scintillation, particularly when the angular scattering is highly asymmetric, one often sees a large number of inverted arcs with their apexes distributed along or near the primary arc \citep{Brisken+10}. We do not see any evidence of inverted arcs in our observations, which are in much weaker scintillation.


The Doppler shift $f_t$ depends on the effective velocity, $V_{\rm eff}$ of the line-of-sight relative to the medium (which is a linear combination of the transverse velocities of the Earth, pulsar, and IISM). With multiple measurements of parabolic arcs at different epochs, the arc curvature will show cyclical variations due to the orbital motions of the Earth and the pulsar. However early analyses of scintillation arcs have primarily involved solitary pulsars without binary motions \citep[e.g.][]{Stinebring+01}, and individual epochs of observations \citep[e.g.][]{Brisken+10, Bhat+16}. One previous example of annual and orbital velocity modulations to arc curvatures has been reported, which was for an analysis of PSR~J0737$-$3039A \citep{Stinebring+05}. However this arc curvature model for PSR~J0737$-$3039A was inferior to an earlier model of scintillation timescale variations observed in the dynamic spectrum \citep{Ransom+04}, because the arcs were not sharp. \citet{Main+20} have also observed annual arc curvature variations in the millisecond pulsar, PSR~J0613$-$0200.

PSR~J0437$-$4715 was the first pulsar to have its full three-dimensional orbital motion determined \citep{vanStraten+01}, and it also has the most precisely measured distance of any pulsar, $D=156.79\pm 0.25$\,pc \citep{Reardon+16}. This makes the system an ideal candidate for modelling of the transverse motion probed by scintillation, and allows us to measure properties of the scattering screens with unprecedented precision. Scintillation arcs have previously been observed for PSR~J0437$-$4715 in observations from the Parkes radio telescope and Murchison Wide-field Array \citep[MWA;][]{Bhat+16, Bhat+18}. However with only two observations of the brightest arc, \citet{Bhat+16} could only estimate the distance to one screen. 

In this paper we show that measuring the scattering screen distance from individual observations in this way, with the necessarily restrictive assumptions of a stationary IISM, can result in significantly biased measurements. Long-term modelling provides a robust way to measure the screen distance, which is constrained by the relative amplitudes of the arc curvature modulation due to the orbits of the pulsar and the Earth. It can also provide precise measurements of the IISM velocity and anisotropy angle, and determine binary orbital parameters such as $i$ and $\Omega$. These pulsar parameters are important for constraining neutron star masses and for tests of general relativity using relativistic binaries, but are difficult to measure through pulsar timing alone. Modelling of the long-term changes to the diffractive scintillations (using the characteristic time- and frequency-scales of the dynamic spectrum) can also be used to precisely measure these parameters \citep[e.g.][]{Rickett+14, Reardon+19}, but the work in this paper is a new approach that can work well even in the regimes of weak and/or time-varying levels of interstellar turbulence.

Our observations make use of the second data release of the PPTA \citep{Kerr+20}, and are briefly described in Section \ref{sec:observations}. In Section \ref{sec:interpreting} we provide theoretical motivation for the arcs in the secondary spectra, discuss anisotropic scattering in the context of our observations, and describe our method for fitting the arc curvature. Section \ref{sec:modelcurve} details the model for the effective velocity and arc curvature variations, and the results are then presented in Section \ref{sec:results}. We identify curvature variations for two arcs corresponding to two discrete scattering screens, and we are able to model the long-term modulations for both. In the discussion in Section \ref{sec:discussion} we give suggestions for candidate structures in the IISM responsible for the scattering and predict how our techniques will be extended and used for interpreting more sensitive observations from new telescopes such as MeerKAT \citep{Bailes+18} and instruments such as the Parkes ultra-wideband receiver \citep{Hobbs+19}.

\section{Observations and data}
\label{sec:observations}

Our observations of PSR~J0437$-$4715 are from the Parkes 64-m radio telescope and span 16\,years from MJD 52618 to 58523 (December 2002 to February 2019), with the majority being taken as part of the PPTA project \citep{Manchester+13} that commenced regular observations in 2005. The PPTA observes a set of millisecond pulsars approximately every three weeks in three observing bands, 40/50-cm (at centre frequencies $f_c\sim 685$\,MHz and $f_c\sim 732$\,MHz respectively), 20-cm ($f_c\sim1369$\,MHz), and 10-cm ($f_c\sim3100$\,MHz). Here we use observations from the 40/50-cm and 20-cm bands because the 10-cm observations do not show clear scintillation arcs in the secondary spectra. Details of the observing systems are described in \citet{Manchester+13} and of the data processing are in \citet{Kerr+20}.

PSR~J0437$-$4715 is highly linearly-polarized and for this reason it is the target of an observing campaign (observing code ``P737"), which tracks the pulsar (a few times per year) for up to $\sim10$\,hours from rise to set for the purpose of polarisation calibration and instrument commissioning. These long observations provide long dynamic spectra and improve the signal-to-noise ratio of any scintillation arcs in the secondary spectra.

A dynamic spectrum from one of these long tracks is shown in the left panel of Figure \ref{fig:specs}, with a well-defined scintillation arc apparent in the secondary spectrum (right panel of Figure \ref{fig:specs}). While these longer observations show particularly clear scintillation arcs, we also detect the primary arc in all available observations in the 20-cm and 50-cm bands, provided they are not too contaminated with radio-frequency interference.

\subsection{Computing dynamic and secondary spectra}
\label{sec:dynspecs}

The dynamic spectra, $S(t,\nu)$, are computed as part of the data processing pipeline that has been developed for the second data release of the PPTA, which uses \textsc{psrflux} from the \textsc{psrchive} package \citep{Hotan+04, vanStraten+12}. This pipeline has already been used to study the relativistic binary pulsar PSR J1141$-$6545 \citep{Reardon+19} to measure scintillation velocity from diffractive scintillation. 

The dynamic spectra have typical resolutions of order $B_c\sim0.5$\,MHz in frequency and $t_{\rm{sub}}\sim30$\,s in time. Data flagged as RFI are replaced using linear interpolation. Removing the RFI reduces artifacts in the secondary spectrum (particularly along the axes) but does not affect curvature measurements. For the long tracks where we concatenate multiple dynamic spectra, the gaps (during which time a noise diode is observed) are filled with the mean flux. Before calculating the secondary spectrum we first re-sample the dynamic spectrum uniformly in wavelength rather than frequency (using cubic interpolation onto a grid with wavelength step size equal to the difference in the lowest two frequency channels), $S(t,\lambda)$, as has been done previously \citep{Fallows+14}. This has the effect of removing the frequency-dependence of the arc curvature and therefore sharpens the features of the arc to improve the curvature measurements.

We also apply a Hamming window function to the outer 10\% of each dynamic spectrum to reduce sidelobe response that adds power along the secondary spectrum axes. We then subtract the mean flux before computing the secondary spectrum, $P(f_t, f_\lambda)$, which is its two-dimensional Fourier transform. This is computed by first pre-whitening the dynamic spectrum (using the first-difference method) before it is Fourier transformed with zero padding. We take the squared magnitude of the transform, and shift and crop it to show only values for $f_\lambda>0$. The spectrum is then ``post-darkened" \citep[the reverse process of pre-whitening][]{Coles+11} and given in units of dB. So in summary we have \begin{equation}
P(f_t, f_\lambda)=10\log_{10}(|\tilde{S}(t,\lambda)|^2),
\end{equation}\noindent where the tilde denotes the two-dimensional Fourier transform, $f_t$ and $f_\lambda$ are the Fourier conjugates of $t$ and $\lambda$ respectively, and $\tilde{S}(t,\lambda)$ is the mean-subtracted and windowed dynamic spectrum with wavelength. A secondary spectrum computed in this way is shown in the right panel of Figure \ref{fig:specs}. 

\subsection{The normalized secondary spectrum 
}
\label{sec:norm_sspec}

\begin{figure*}
\centering
\includegraphics[width=.32\textwidth]{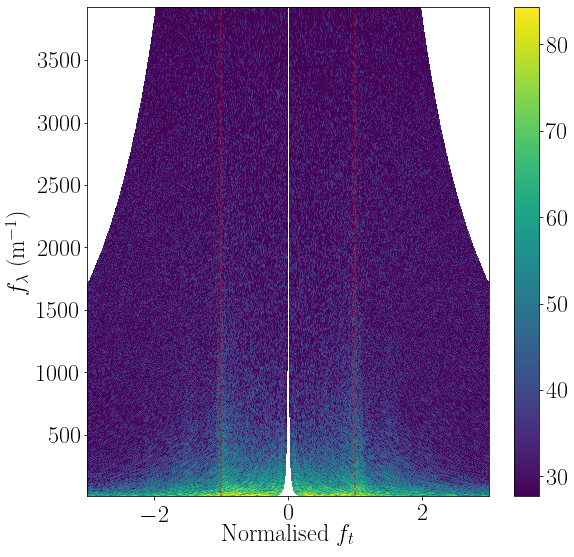}
\includegraphics[width=.32\textwidth]{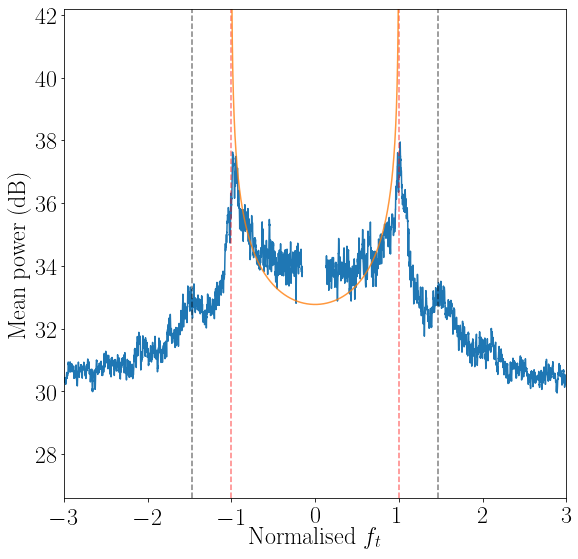}
\includegraphics[width=.32\textwidth]{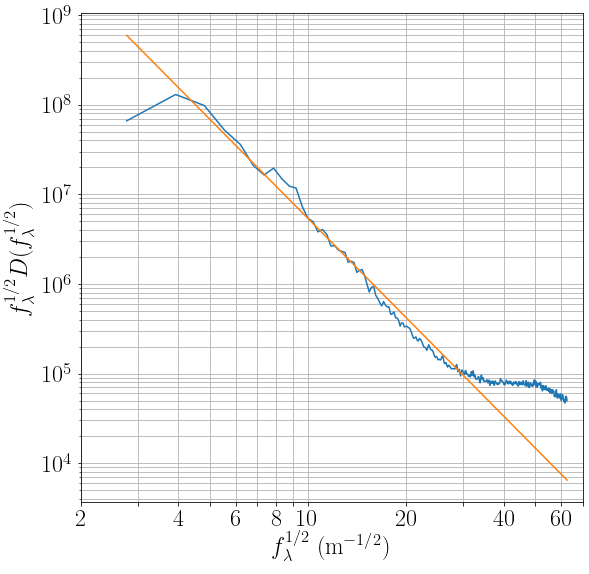}
\caption{The normalized secondary spectrum $P(f_t/f_{arc}, f_\lambda)$ (left) of the observations shown in Figure \ref{fig:specs}. The normalizing $f_{arc}$ is found by a best fit for the curvature of the primary arc. We refer to the marginal distributions as the Doppler profile (center) and the Delay profile (right). The Doppler profile $D_t (f_t/f_{arc})$ shows how the power decays inside the arc, which is related to the strength of scintillation and the anisotropy. The secondary arc can be seen in these spectra, at normalised $f_t$, $f_{tn} \sim 1.5$ (indicated by the black dashed lines). The Delay profile, shown as $f_\lambda^{1/2} \ D_\lambda ({f_\lambda^{1/2}})$, follows the phase spectrum. The orange curve on the marginal distributions shows the weak scintillation approximation for an isotropic image from Kolmogorov turbulence (Equation \ref{eqn:Doppler_prof}).}
\label{fig:norm_sspec}
\end{figure*}

We introduce a novel way to search for forward arcs in secondary spectra and analyse their power distributions, by re-sampling the spectrum in Doppler to transform any parabolas into vertical lines. This is done by adjusting the sampling for each row of the spectrum, with linear interpolation, such that the number of samples decreases with $f_\lambda^2$. We refer to this a ``normalized" secondary spectrum, $P(f_t/f_{arc}, f_\lambda)$, with respect to some reference arc curvature $\eta$, when the transformation is done such that the units on the x-axis are $f_{tn} = f_t / f_{arc}$, the fractional distance from the $f_t=0$ axis to the arc at $f_{arc}$, for a given $f_\lambda$. In this way, any arc at $\eta$ will become a vertical line of power at the normalized $f_{tn}=1$, and for example, a second vertical line of power at normalized $f_{tn}=\beta$ would correspond to a second arc with curvature $\eta_{\beta} = \eta/\beta^2$. The normalized secondary spectrum for the data shown in Figure \ref{fig:specs}, is given in the left panel of Figure \ref{fig:norm_sspec}.

Taking cuts across $P(f_{tn}, f_\lambda)$ along the $f_\lambda$ axis, shows that the shape of the power distribution is approximately constant with $f_\lambda$, however the amplitude decays steeply as $\sim f_\lambda^{-7/3}$ (see Appendix B). This constant profile shape means that $P(f_{tn}, f_\lambda)$ can be averaged over $f_\lambda$ (with appropriate weighting) to increase the signal-to-noise ratio of the power distribution across and inside the arc. This Doppler profile $D_t(f_{tn})$ is shown in the center panel of Figure \ref{fig:norm_sspec}. It is useful for analysing the anisotropy of the scattering and for fitting arc curvatures.
The Delay profile $D_\lambda(f_\lambda)$ is obtained by averaging over Doppler. In weak scintillation the Delay profile is power-law with an exponent of $\alpha/2 - 1$ where $\alpha$ is the spectral exponent of the turbulence (see Appendix B). Accordingly we display the Delay profile in the right panel of Figure \ref{fig:norm_sspec} scaled by ${f_\lambda^{1/2}}$ and plotted vs ${f_\lambda^{1/2}}$. The resulting plot is directly proportional to the phase spectrum in weak scintillation, if Kolmogorov the spectral exponent would be -11/3.

\section{Interpreting and fitting the secondary spectra}
\label{sec:interpreting} 

In this Section we give a brief overview of scintillation arcs, with a focus on the curvature parameter and a discussion relevant to the expected Doppler profiles of the secondary spectra. Detailed reviews of the theory of scintillation arcs can be found in \citet{Walker+04} and \citet{Cordes+06}.

A power-law distribution of density irregularities in the IISM scatters incident radiation, by means of diffraction, into a spectrum of angles relative to the direct line-of-sight to the source. The interference of waves arriving at the observatory from two small angles in this spectrum, 
$\vec{\theta}_1$ and $\vec{\theta}_2$, 
produces a single frequency-dependent interference fringe pattern, which is observed in time and frequency as a sinusoid in the dynamic spectrum \citep{Cordes+06}. For a geometrically thin (in the radial direction) screen at some fractional position $s$ along the line of sight from the source (where $s=0$ is at the source and $s=1$ is at observatory), the axes of the wavelength-resampled secondary spectrum $P(f_t, f_\lambda)$ are related to these scattering angles with
\begin{equation}
\label{eqn:tdel}
f_\lambda = \frac{D(1-s)}{2s\lambda_c^2}(\theta_2^2 - \theta_1^2)
\end{equation}\begin{equation}
f_t = \frac{1}{s\lambda_c}\vec{V}_{\rm{eff}}\cdot(\vec{\theta}_2-\vec{\theta}_1),
\label{eqn:fdop}
\end{equation}

\noindent where $D$ is the distance to the source from the observatory, $\vec{V}_{\rm{eff}}$ is the effective velocity of the line-of-sight through the screen (Equation \ref{eqn:veff1}), $\lambda_c$ is the central wavelength of the observation, and $c$ is the speed of light. The Fourier variable conjugate to $\lambda$,  $f_\lambda = c\tau_{\rm{del}}/{\lambda_c}^2$, where $c$ is the speed of light, $\tau_{\rm{del}} = f_\nu$ is the differential geometric time delay between the paths taken to arrive from the two angles, and $f_t$ is their differential Doppler shift. Each Fourier component in the secondary spectrum corresponds to a particular sinusoidal fringe pattern and thus to the summation of all pairs of components of the angular spectrum with the same values of $f_\lambda$ and $f_t$.

\begin{figure*}
\centering
\includegraphics[width=.95\textwidth]{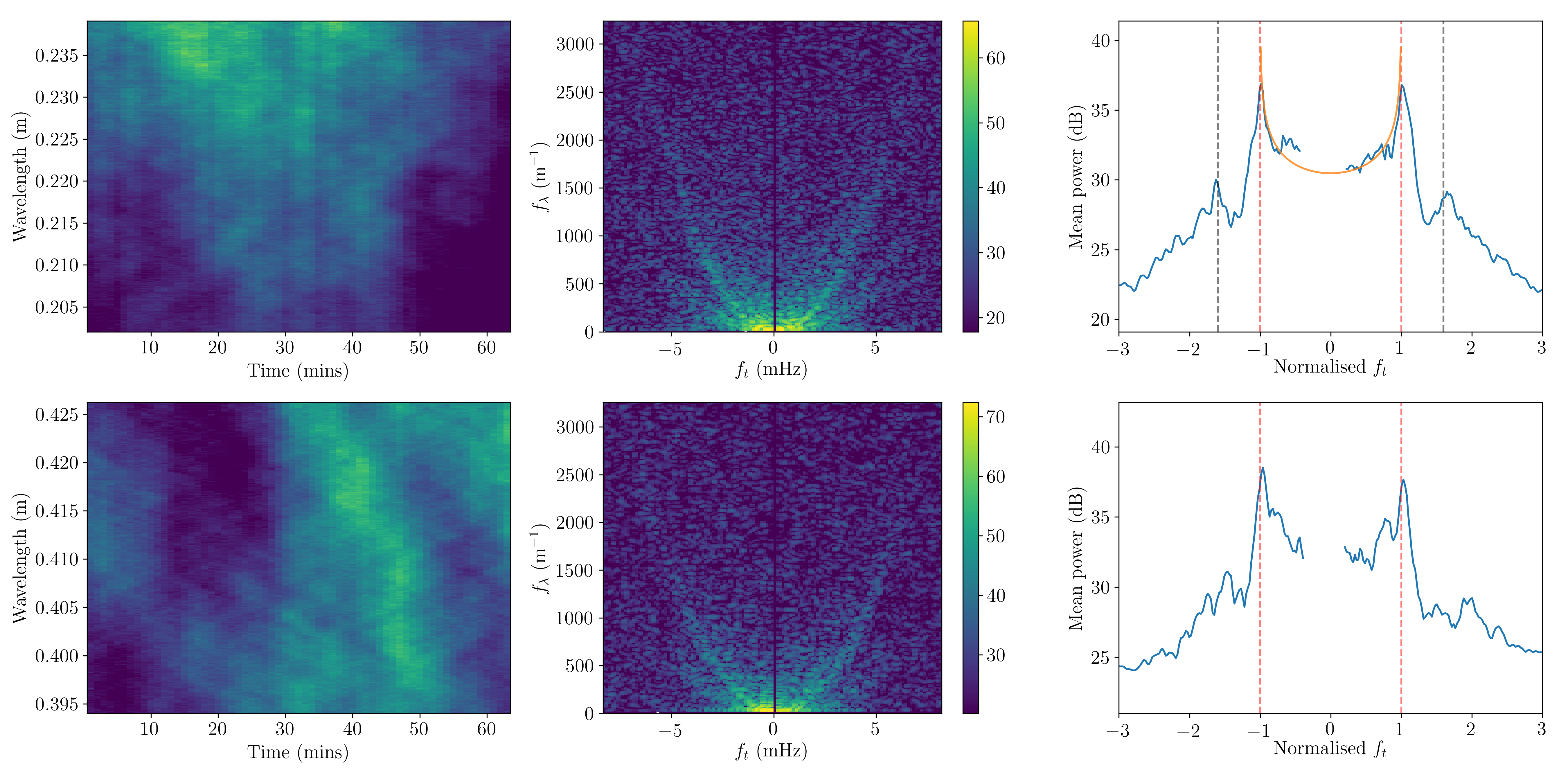}
\caption{Wavelength-resampled dynamic spectra (left panels), secondary spectra (middle panels), and Doppler profiles (right panels, with x-axis normalised with respect to the primary arc) for a typical 20-cm observation (on MJD 55832; top) and 40-cm observation (on MJD 56319; bottom). The 20-cm observation was included in our dataset of curvature measurements for the secondary arc, and the measured value is indicated by the black dashed lines. The orange curve on the 20-cm Doppler profile shows the expected Doppler profile for an isotropic image in the weak scintillation regime (Equation \ref{eqn:Doppler_prof}).}
\label{fig:example_obs}
\end{figure*}

The form of the secondary spectra depends strongly on the strength of scintillation. This is defined by the phase structure function $D_\phi(s) = (s/s_0)^\alpha = \langle (\phi(r) - \phi(r+s))^2\rangle$. The intensity variance in weak scintillation $m_b^2$ is the accepted measure of the strength of scintillation. For an isotropic Kolmogorov spectrum \citep[Equation 3 of][]{Coles+10}  $m_b^2 = 0.773\  D_\phi(r_F)$ where $r_F$ is the Fresnel scale. 
This is most easily measured by the fractional bandwidth of the  diffractive scintillations, which is given by $\Delta\nu_d/\nu_c = (s_0/r_F)^2$ \citep{Rickett90}. For a Kolmogorov structure function, $D_\phi (r_F) = (r_F/s_0)^{5/3}$, and using this we can estimate the Born variance as $m_b^2 = 0.773\  (\nu_c/\Delta\nu_d)^{5/6}$. 
\noindent In this way we find $m_b^2 \approx 4$ for the 20-cm observations and $m_b^2 \approx 22$ at 40-cm. These values are appropriate for simulating scintillation with our observed $\Delta\nu_d/\nu_c$ under the assumption of isotropic scattering. However if the scattering is anisotropic then $\Delta\nu_d/\nu_c$ will be reduced \citep{Rickett+14}. To account for this in simulations, we iteratively decrease $m_b^2$ until the simulated $\Delta\nu_d/\nu_c$ is within 5\% of the isotropic simulation.

We use the technique described in \citep{Coles+10} to simulate dynamic spectra with the same fractional bandwidth of our observations. In this we implicitly assume that the scintillation is dominated by the thin screen which provides the primary arc. We have reproduced this technique in Python and have made it publicly available\footnote{From \url{https://github.com/danielreardon/scintools}. See also Appendix A}. 


In weak scintillation, where $m_b^2 < 1$, the parabolic arc in the secondary spectrum can be interpreted as the interference between the unscattered image of the pulsar and the surrounding scattered image. In this case we have Equations \ref{eqn:tdel} and \ref{eqn:fdop} with $\theta_1=0$ and $\theta_2=\theta_0$, where $\theta_0$ is the angular separation between a component of the scattered image and the line-of-sight. The arc curvature then comes from the quadratic relationship between $f_\lambda$ and $f_t$ through their dependence on $\theta_0$. We introduce the curvature parameter, $\eta$, such that  $f_\lambda = \eta f_t^2$, and this is then given by

\begin{equation}
\label{eqn:eta}
\eta=\frac{Ds(1-s)}{2V_{\rm{eff}}^2\cos^2\psi}
\end{equation}

 \noindent where $\psi$ is the angle between $V_{\rm{eff}}$ and the position vector along the anisotropy in the scattered image and comes from the dot product in Equation \ref{eqn:fdop}. For an isotropically-distributed image, such as a ring or halo, the equation is the same, but with $\cos\psi=1$ describing the outer-edge of the arc \citep{Cordes+06}. We give the equation for Doppler profiles in weak scintillation in Appendix B, Equation \ref{eqn:Doppler_prof}.

In the case of strong scintillation (such as in 40/50-cm observations of PSR~J0437$-$4715), the scattered image that extends beyond the root-mean-square (rms) scattering angle interferes with itself as well as the main image. Anisotropic scattering in this regime leads to inverted parabolas referred to as arclets \citep[e.g.~][see also Figures \ref{fig:anisotropy_sims} and \ref{fig:orientation_sims} in Appendix B for secondary spectra containing arclets]{Brisken+10} with their apexes distributed along the main arc, unless they appear at different angles $\psi$ with respect to the position angle of $\vec{V}_{\rm{eff}}$. In the case of bright, discrete scattered images beyond the rms scattering angle, these arclets can be resolved individually \citep[as in][]{Brisken+10}. However, more continuous anisotropy at high scattering angles is expected to produce a forest of these arclets that can together appear as a broadened scintillation arc \citep[see examples in][]{Cordes+06}. In this regime, the curvature described by Equation \ref{eqn:eta} is the line through the centres of the arclets, rather than the outer edge of power.

Arcs may also show asymmetries in their total power about $f_t=0$ (Doppler asymmetries) formed by asymmetric scattering about the line-of-sight in the direction of $\vec{V}_{\rm{eff}}$ \citep[e.g.][]{Cordes+06}. This can be caused by a phase (and electron density) gradient across the line-of-sight, or physically asymmetric structures in the scattering medium. We do not observe any clearly asymmetric arcs at any epoch or orientation of the $\vec{V}_{\rm{eff}}$ vector. This may indicate that the scattered image is symmetric, such as a linear structure (for the case of highly anisotropic scattering), an ellipse (for a moderately anisotropic image), or a circularly-symmetric halo (for isotropic scattering).

\subsection{Fitting arc curvature}
\label{sec:fitarcs}

To measure the arc curvature and estimate its uncertainty from the secondary spectrum of each observation, we use the Doppler profiles described in Section \ref{sec:norm_sspec}. We re-scale the x-axis of the normalized secondary spectrum into physical units of the curvature, $\eta$. The transformation to this Doppler profile is computed only once for each observation and displays the mean power as a function of arc curvature $P(\eta)$, so curvature measurements are found simply by detecting peaks in this distribution.

\begin{figure}
\centering
\includegraphics[width=.4\textwidth]{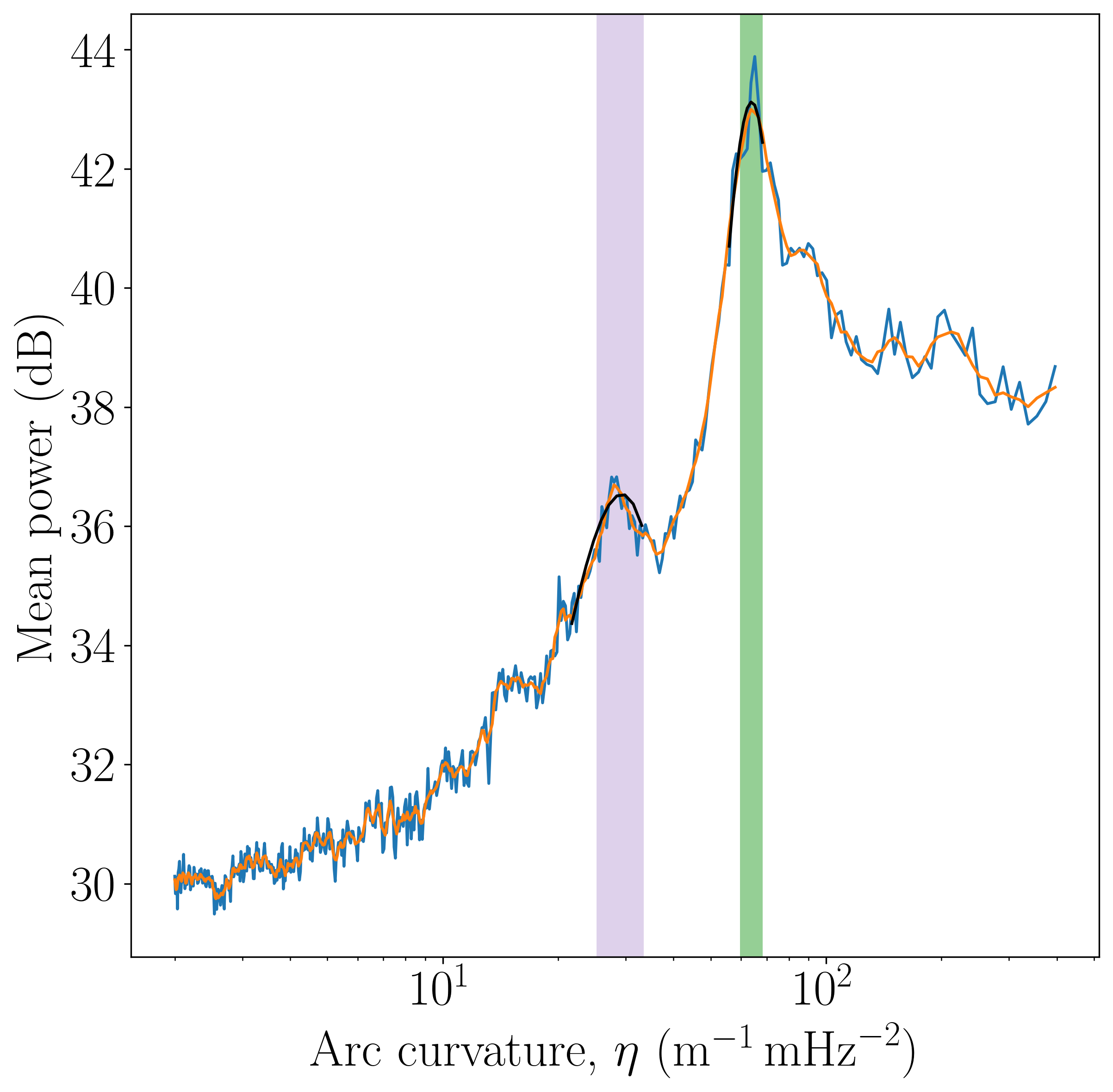}
\caption{Result of arc curvature fitting to the secondary spectrum Doppler profile shown in Figure \ref{fig:norm_sspec}. Here the positive and negative sides of the Doppler profile have been averaged to increase the signal-to-noise ratio. The measurement and uncertainty regions for the primary and secondary arcs are shown with the green and purple bands respectively. The orange line is the smoothed data that were used to select the fitting regions, and the inverted parabolas in black show the result of these fits. The arc curvature axis is displayed on a logarithmic scale for visualization only.}
\label{fig:curve_fit}
\end{figure}

The $P(\eta)$ curve for our long 20-cm observation from Figures \ref{fig:specs} and \ref{fig:norm_sspec} is shown in Figure \ref{fig:curve_fit}. The primary and secondary arcs are clearly seen as peaks in $P(\eta)$. For most of our observations, the signal-to-noise of the arc peak is not as high as in this figure, because most observations have a 64\,min duration. Some more typical examples from 20-cm and 40-cm observations are shown in Figure \ref{fig:example_obs}. The curvature measurements for peaks in this distribution are found by first smoothing\footnote{Using a first-order Savitzky-Golay filter.} the data, and then fitting a parabola to the un-smoothed data in a region from $-3$\,dB on the low-curvature side to $-1.5$\,dB on the high-curvature side around the peak in the smoothed data. We choose an asymmetric window to fit the data because the power drops off more steeply on the low-curvature side than the high-curvature side. This asymmetric window fits closely to the arc's sharp outer edge and minimises the effect of additional power inside the arc. 

The uncertainty on our curvature measurement  $\sigma_\eta$ is determined from the noise level in the secondary spectrum, far from the power in the arc $\sigma_s$ \citep[as in][]{Bhat+16}. We construct a standard error confidence region around the curvature measurement, corresponding to the change in $\eta$ required for the power to drop by $\sigma_s$, from the peak in the smoothed data. The  curvature measurement and the uncertainty regions for the primary and secondary arcs are shown in Figure \ref{fig:curve_fit}.

Using this method, the secondary spectrum must be cropped (or truncated) at a $\tau_{\rm{del}}$ (or its corresponding $f_\lambda$ after wavelength re-sampling) value that is just beyond where the arc power becomes lower than the noise. This is difficult because the power in the arcs decays with increasing $\tau_{\rm{del}}$. We chose to crop each secondary spectrum at a fixed maximum time delay, $\tau_{\rm{del},max}$ beyond which most observations show no evidence of the primary arc. This delay depends on the observing frequency for the observation, and we have therefore defined $\tau_{\rm{del},max} = 0.25\,\mu s \times (1400\,{\rm MHz}/f)^2$ for the primary arc and $\tau_{\rm{del},max} = 0.1\,\mu s \times (1400\,{\rm MHz}/f)^2$ for fitting the secondary. In general this is a conservative figure that will include noise in our data and our estimated uncertainties for the arc curvature are expected to be slightly overestimated as a result. 

Figure \ref{fig:curve_fit} also shows that there are potentially low signal-to-noise arcs at $\eta \approx 13$\,m$^{-1}$\,mHz$^{-2}$, and $\eta \approx 200$\,m$^{-1}$\,mHz$^{-2}$. We do not analyse these arcs in detail because we cannot reliably measure their curvatures for a significant number of epochs. They appeared at multiple epochs only in the long track observations, and always with a low signal-to-noise ratio.

\section{Modelling arc curvature variations}
\label{sec:modelcurve} 

The curvature of arcs in the secondary spectrum depends on the distance to the scattering region $s$ (assumed to be a thin screen), and the velocity of the line-of-sight with respect to the medium, $\vec{V}_{\rm{eff}}$, as given in Equation \ref{eqn:eta}. We therefore expect the curvature to be time-dependent as $\vec{V}_{\rm{eff}}$ changes because of the changing transverse components of the Earth's velocity ($\vec{V}_{\rm{E}}$) and the pulsar's orbital velocity ($\vec{V}_{\rm{p}}$). The effective velocity is a linear combination of these velocities and the velocity of the medium itself ($\vec{V}_{\rm{IISM}}$) \citep{Cordes+98},
\begin{equation}
\label{eqn:veff1}
\vec{V}_{\rm{eff}} = (1 - s)(\vec{V}_{\rm p} + \vec{V}_{\mu}) + s\vec{V}_{\rm E} - \vec{V}_{\rm{IISM}},
\end{equation}
\noindent where $\vec{V}_{\mu}$ is the constant transverse velocity of the pulsar system (corresponding to its proper motion).

We model the variations in $\eta(t)$ with $\vec{V}_{\rm{eff}}$ components in right ascension ($\alpha$) and declination ($\delta$)
\begin{equation}
\label{eqn:veff2}
\begin{aligned}
v_{\rm{eff},\alpha} =& (1 - s)(v_{\rm p,\alpha} + v_{\rm \mu,\alpha}) + sv_{\rm E,\alpha} - v_{\rm{IISM},\alpha}\\ 
v_{\rm{eff},\delta} =& (1 - s)(v_{\rm p,\delta} + v_{\rm \mu,\delta}) + sv_{\rm E,\delta} - v_{\rm{IISM},\delta}\\ 
\vec{V}_{\rm{eff}}=&\sqrt{v_{\rm{eff},\alpha}^2+v_{\rm{eff},\delta}^2}
\end{aligned}
\end{equation}The distance and proper motion for PSR~J0437$-$4715 are known to high precision from pulsar timing, giving $v_{\rm \mu,\alpha}=90.25\pm 0.15$\,km\,s$^{-1}$ and $v_{\rm \mu,\delta}= -53.12 \pm 0.09$\,km\,s$^{-1}$ \citep{Reardon+16}. The precise timing model also allows us to derive the mean orbital velocity 
\begin{equation}
V_0=\frac{2\pi xc}{\sin{i}P_b \sqrt{\left(1 - e^2\right)}}=18.946\pm0.015\,{\rm km\,s}^{-1}
\end{equation}
\noindent from the projected semi-major axis $x$ (in light-seconds), orbital period $P_b$, eccentricity $e$, and the inclination angle $i$. The orbital transverse velocity in components parallel and perpendicular to the line of nodes ($v_{\rm p,\parallel}$, and $v_{\rm p,\perp}$ respectively) is then defined in terms of the true orbital anomaly $\theta$
\begin{equation}
\label{eqn:vcomponents}
\begin{aligned}
v_{\rm p,\parallel} &= -V_0\left(e\sin{\omega} + \sin{(\theta+\omega)}\right)\\
v_{\rm p,\perp} &= V_0\cos{i}\left(e\cos{\omega} + \cos{(\theta+\omega)}\right),
\end{aligned}
\end{equation}

\noindent where $\omega$ is the longitude of periastron. These components are rotated into right ascension $\alpha$ and declination $\delta$ with the longitude of the ascending node $\Omega$, defined East of North. This is the most uncertain parameter in the pulsar timing model, with $\Omega=207.0\pm 1.2^\circ$, because it is measured through a subtle kinematic effect caused by the combination of the Earth's orbital motion and pulsar's transverse velocity, which changes the projection of the orbit \citep{Kopeikin95}. 

Remarkably, by modelling the annual and orbital modulation of diffractive scintillations \citep[e.g.][]{Rickett+14, Reardon+19} or arc curvature, we are able to measure $\Omega$ and other parameters (such as $i$) often with higher precision than through pulsar timing, because these parameters have a strong influence on the transverse velocity variations. From Equations \ref{eqn:veff2} and \ref{eqn:vcomponents} we see that the only parameters required for modelling the arc curvatures for PSR J0437$-$4715 are: $s$, $v_{\rm{IISM},\alpha}$, and $v_{\rm{IISM},\delta}$, although we also fit for $\Omega$ and $i$, as consistency checks for our model. Our measurements of these parameters therefore demonstrate the precision and accuracy achievable from the scintillation arcs, which may be useful for other systems that do not have such precise timing. 

The models are fitted to the data using the \textsc{emcee} \citep{Foreman-Mackey+13} Markov chain Monte Carlo algorithm within the \textsc{lmfit} python package \citep{Newville+14}. This allows us to probe the full posterior probability distribution of our model, derive robust measurement of the parameter uncertainty, and visualize any parameter correlations.

\section{Results}
\label{sec:results} 

\begin{figure*}
\centering
\includegraphics[width=.9\textwidth]{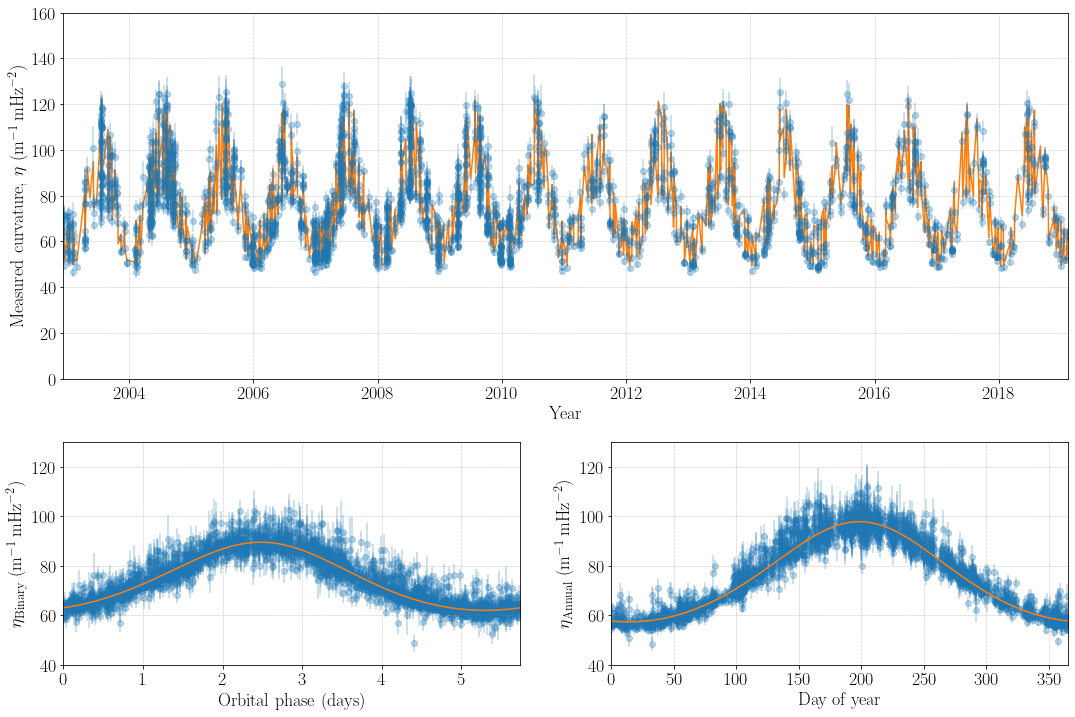}
\caption{Arc curvature measurements for the primary scintillation arc (top panel) with the best-fit model shown as the orange line. The annual cycle component of this model was subtracted from the data and model to show the orbital modulation with $\sim 5.7$\,day period in the lower left panel, while the orbital cycle was subtracted to show the annual modulation in the lower right panel.}
\label{fig:primary_curvature}
\end{figure*}

We have measured the curvature of scintillation arcs in a set of observations across two observing bands for PSR~J0437$-$4715. We have found that all observations in the 20-cm and 40/50-cm bands show evidence for at least one arc, the ``primary'', which is the strongest in all of the observations. The arc signal-to-noise ratio depends strongly on the pulsar flux, observation length $t_{\rm obs}$, and observing bandwidth $B$. In observations with the highest signal-to-noise ratio, we see a fainter, secondary arc at a lower curvature.

The time series of $\eta$ for each measurement of the primary scintillation arc is shown in Figure \ref{fig:primary_curvature}. There is a clear annual modulation to the curvatures, as well as a $\sim5.7$\,day modulation corresponding to the pulsar's orbital period. We find similar annual and binary orbital modulation for the secondary arc, although the binary orbital modulation dominates for this arc because the scattering region is located closer to the pulsar on the line-of-sight. Table \ref{tab:Params} gives the fitted parameters and $\chi^2$ values for the models of each of these arcs, and we describe these models in detail in the following sections.

\subsection{Testing for anisotropy}

Extreme anisotropy of the scattered image cannot be assumed a priori. Axial ratios of order 2 have been observed in the two relativistic binaries whose scattering geometry and three-dimensional orbits have been completely solved \citep{Rickett+14, Reardon+19}. However the solitary source PSR~B0834+06, which has also been solved with the help of VLBI scintillations \citep{Brisken+10}, shows extremely anisotropic scattering. For this work we fit models for both isotropic and anisotropic scattering to both of our measured arcs. 

\begin{figure*}
\centering
\includegraphics[width=.4\textwidth]{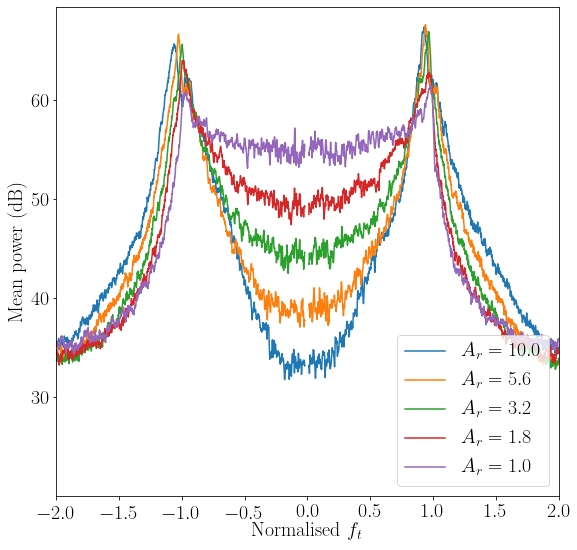}
\includegraphics[width=.4\textwidth]{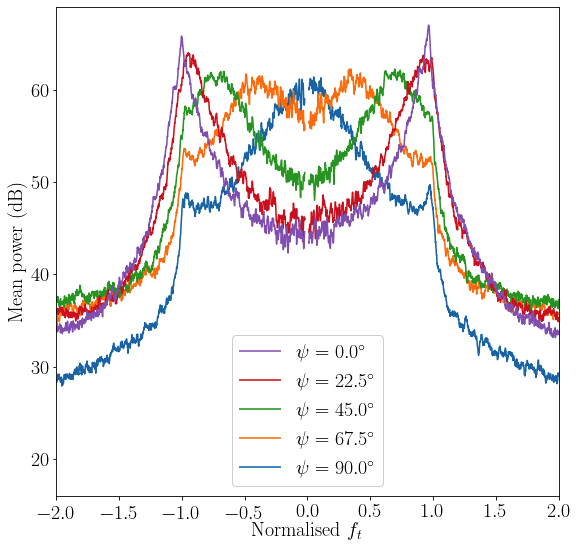}
\caption{Simulated Doppler profiles for secondary spectra with $m_b^2$ selected to match the observed fractional scintillation bandwidth for 20-cm observations of PSR~J0437$-$4715. The left panel shows the effect of varying the degree of anisotropy for a fixed orientation (in line with the velocity vector), while the right panel shows the effect of varying orientation for a fixed anisotropy ($A_r=3.2$). The full simulated secondary spectra and the separate Doppler profiles for each of these curves are given in Appendix B, Figures \ref{fig:anisotropy_sims} (for the left panel) and \ref{fig:orientation_sims} (for the right panel).}
\label{fig:Anisotropy_simulation}
\end{figure*}

In addition to a goodness-of-fit assessment, we use existing theory for the expected power distribution in the secondary spectra to give us some additional insight into the validity of these models. Fortunately we have both 20-cm and 40-cm observations. The latter have a scintillation bandwidth of $\Delta\nu_{\rm d}\sim$25 MHz, which translates to $\sim$400 MHz at 20-cm, following a $\Delta\nu_d \propto f^{-4}$ relationship. While the 40-cm observations are in the strong scintillation regime, the 20-cm observations are in the transition between weak and strong. For this transition scattering the theory is not well-developed, but we can simulate the scattering using the techniques discussed by \citet{Coles+10}. The effect of anisotropy on weak scintillation is best illustrated in the Doppler profiles. Examples of such simulated profiles, for a strength of scintillation that matches our observations in the 20-cm band, are shown in Figure \ref{fig:Anisotropy_simulation}. The full secondary spectrum corresponding to these profiles, as well as the analytical profile for the case of weak scintillation, are given in Appendix B.

\begin{table*}
\centering
\caption{Parameters from models of the curvature variations in the primary and secondary arcs for PSR~J0437$-$4715. The values in brackets are the uncertainty on the last quoted decimal place. The full posterior probability distributions for the primary arc models are shown in Appendix B in Figure \ref{fig:posterior}. For comparison, the parameters measured independently from pulsar timing are i = 137.56(4)$^{\circ}$ and $\Omega = 207.0(12)^{\circ}$ \citep{Reardon+16}. For the secondary arc, we also present models with the inclination angle fixed at this timing value, which significantly improves the precision for $s$. The goodness of fit is quantified with the $\chi^2$ value for each model.}
\begin{tabular}{lllllll}
\hline
\hline
& \multicolumn{2}{c}{Primary} & \multicolumn{4}{c}{Secondary} \\
& \multicolumn{1}{c}{Isotropic} & \multicolumn{1}{c}{Anisotropic}  & \multicolumn{2}{c}{Isotropic} &  \multicolumn{2}{c}{Anisotropic}   \\
&  & &  Fitted $i$ & Fixed $i$ & Fitted $i$ & Fixed $i$\\
\hline
$s$   &  $0.424(2)$ &  $0.427(2)$  &  $0.22(4)$ & $0.212(16)$   &  $0.21(4)$ &  $0.215(16)$ \\
$i$ ($^\circ$)& $136.9(3)$    &  $137.1(3)$  & $138(4)$ & \nodata   & $137(4)$  & \nodata  \\
$\Omega$ ($^\circ$, N$ \rightarrow$ E)  &   $205.9(4)$   & $206.3(4)$   &  $211(6)$ & $211(6)$    & $214(6)$ &  $214(6)$ \\
$v_{\rm{IISM},\alpha}$ (km\,s$^{-1}$)&  $-12.0(4)$ & \nodata  &  $-6(11)$  & $-5(10)$  & \nodata & \nodata \\
$v_{\rm{IISM},\delta}$ (km\,s$^{-1}$)&  $31.8(4)$  & \nodata    & $56(10)$   & $56(8)$  & \nodata & \nodata\\
$\xi$ ($^\circ$, N$ \rightarrow$ E) & \nodata & $134.6(3)$  & \nodata & \nodata  &  $144(6)$ & $144(6)$ \\
$v_{\rm{IISM},\xi}$ (km\,s$^{-1}$) & \nodata &  $-31.9(3)$ & \nodata & \nodata    & $-50(10)$ & $-50(6)$  \\
\hline
$\chi^2$ & 2959 & {\bf 2640}  &  158  & {\bf 156}   & 162 & 163 \\
\hline
\end{tabular}
\label{tab:Params}
\end{table*}

The Doppler profile in Figure \ref{fig:norm_sspec} (middle panel) shows a sharp outer edge and a decay in power inside the arc of $\sim 4$\,dB, to a level that is significantly higher than the noise. From Figure \ref{fig:Anisotropy_simulation}, perfectly isotropic scattering in near-weak scintillation results in a sharp outer edge and a decay of $\sim 5$\,dB to the minimum power inside the arc, while larger axial ratios produce much deeper wells of power, for example $\sim 25$\,dB at $A_r \sim 5$. The observed level of power inside the arc is actually slightly higher than that expected from a single isotropic scattering screen, which suggests that there are potentially contributions from other screens. One screen inside this arc is resolved in a few epochs (Section \ref{sec:secondary}), and there may be more unresolved screens.

For the near-weak scintillation at 20-cm, the arcs remain sharp for moderate axial ratios, but at $A_r \gtrsim 5$ inverted arclets begin to appear, which can cause broadening of the arc in the Doppler profile, as well as multiple apparent peaks. These arclets are more pronounced in the stronger scintillation regime of our 40-cm observations. While the quality of our data is poorer in the 40-cm band, the arcs appear to remain sharp (as in Figure \ref{fig:example_obs}), which disfavours extreme anisotropy assuming our strength of scintillation estimate is accurate. The Doppler profiles for our data are well described by small-to-moderate axial ratios $A_r \lesssim 5$. 

If there is any persistent anisotropy, then the curvature of the arcs (measured as the peak power) will be modulated by the $1/\cos^2\psi$ term in Equation \ref{eqn:eta} as the velocity vector moves with respect to the major axis of the anisotropy. This modulation is not exclusive to extreme anisotropy, as demonstrated in the simulations with $A_r = 3.2$ and varying position angle $\psi$ in the right panel of Figure \ref{fig:Anisotropy_simulation}. The transverse velocity for PSR~J0437$-$4715 does not vary in position angle by more than $30^\circ$ on the sky for the primary arc, because it is dominated by the pulsar's proper motion. This is not sufficient to estimate the direction of the anisotropy accurately, but the observed sharpness of the arcs implies that the major axis of any moderate anisotropy must be aligned roughly with the velocity vector.

\subsection{Primary arc}
\label{sec:primary}

The primary arc appears in all observations in the 20-cm and 40/50-cm bands that are not too contaminated with RFI, giving us 2645 unique measurements of $\eta$. The maximum likelihood model for the variations in $\eta$ is shown in Figure \ref{fig:primary_curvature}. 

This model gives a precise measurement of the inclination angle, $i=137.1\pm0.3^\circ$, which differs from the timing solution of $i=137.56\pm0.04^\circ$ \citep{Reardon+16} by $\sim 1.5\sigma$. Similarly, our measurement of the longitude of ascending node $\Omega = 206.3 \pm 0.4^\circ$, is within $\sim 1\sigma$ of the timing measurement $\Omega = 207.0 \pm 1.2^\circ$, and actually surpasses its precision despite this being one of the most precisely timed millisecond pulsars. The precision (and potential accuracy) of these measurements is impressive for scintillation studies, which can often be complicated by changes to properties of the scattering with time. 

The IISM velocity and anisotropy in the direction of PSR~J0437$-$4715 remains stable over the $\sim$16\,years of our observations, meaning that the scattering geometry and kinematics of the screen are stable over a spatial scale of hundreds of AU. Since the proper motion for this pulsar is known to high precision and included in our model, the measured constant components of the velocity are only due to the velocity of the IISM, with magnitude $|V_{\rm IISM}|\gtrsim 32$\,km\,s$^{-1}$. We know very little about the IISM features causing scintillation. This velocity is high compared with the expected thermal or Alfv\'{e}n velocity of the IISM \citep{Goldreich+95}, $\sim 10$\,kms$^{-1}$ , but an entire cloud or outflow could be moving at this rate without any internal disturbance.

This best-fit model is of anisotropic scattering, in which the scattered image is elongated on the sky with axial ratio $A_r \gtrsim 2$. In this model we fit for the orientation of the image on the sky, $\xi$ (defined East of North) and then derive for each observation the angle, $\psi$, between the image and the effective velocity vector. We also fit for the component of the IISM velocity along the image, but not the perpendicular component because the data are completely insensitive to any motion perpendicular to the image. 
As discussed in the previous Section, any anisotropy must be roughly aligned with the velocity to produce arcs that are sharp at all epochs. We find $v_{\rm{IISM},\xi} = -31.9 \pm 0.3$\,km\,s$^{-1}$ and $\xi = 134.6 \pm 0.3^\circ$, which is indeed close to direction of proper motion. 

The measured parameters from the isotropic model are also shown in Table \ref{tab:Params}, and demonstrate that the orbital parameters $i$ and $\Omega$ are nearly independent of the choice of scattering model (isotropic or anisotropic). The distance to the screen $s$ is also nearly model-independent because it is constrained by the relative amplitudes of the annual and orbital modulations. Therefore this technique can be a powerful tool for solving pulsar orbits and accurately determining screen distances.

\subsection{Secondary and additional arcs}
\label{sec:secondary}

\begin{figure*}
\centering
\includegraphics[width=.9\textwidth]{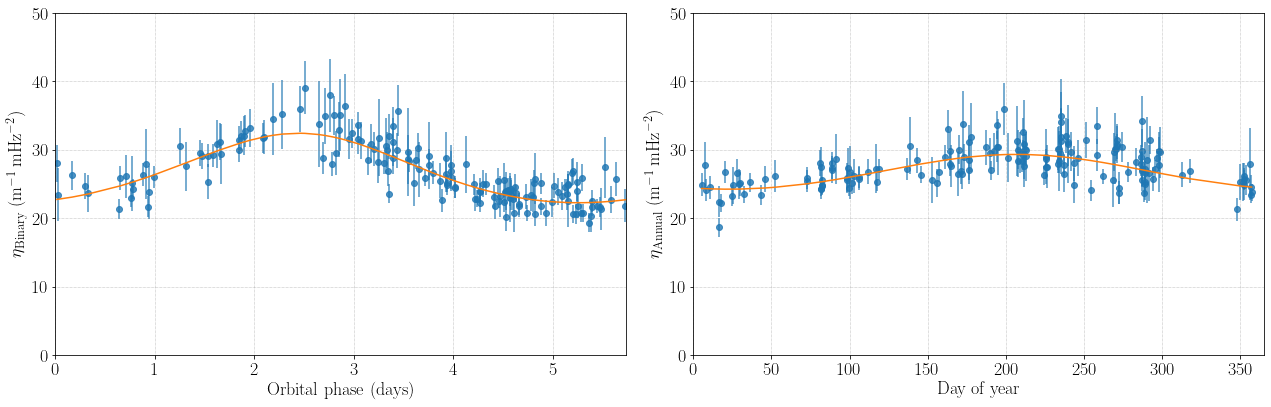}
\caption{Orbital (left panel) and annual (right panel) variations in arc curvature measurements for the secondary arc (as in Figure \ref{fig:primary_curvature} for the primary arc). The isotropic model is shown as an orange line.}
\label{fig:secondary_curvature}
\end{figure*}

We measure the secondary arc using 165 of the highest signal-to-noise ratio observations from the PDFB4 signal processing system. As with the primary arc, we have fitted both isotropic and anisotropic models, with measured parameters given in Table \ref{tab:Params}. In this case the isotropic model provides a slightly better fit to the data. The orbital and annual components of the velocity model for this secondary screen are shown in Figure \ref{fig:secondary_curvature}. 

Using this secondary screen alone, we are able to measure the inclination angle $i=138 \pm 4 ^\circ$ and longitude of ascending node $\Omega=211 \pm 6 ^\circ$ for the isotropic scattering model (with $i=137 \pm 4 ^\circ$ and $\Omega=214 \pm 6 ^\circ$ for the anisotropic model). By fixing $i$ at the superior measurement from pulsar timing, we determine the fractional screen distance to be $s = 0.212 \pm 0.016$. Both of our velocity models for this screen suggest a relatively high IISM velocity of $|V_{\rm IISM}| \sim 50$\,km\,s$^{-1}$.

In approximately half of the long track observations we see evidence for additional arcs. The two most prominent appear in Figure \ref{fig:curve_fit} at curvatures $\eta \approx 13$\,m$^{-1}$\,mHz$^{-2}$, and $\eta \approx 190$\,m$^{-1}$\,mHz$^{-2}$, corresponding to fractional screen distances of approximately $s=0.06$, and $s=0.8$ respectively, under the assumption of a stationary IISM and isotropic scattering. We were unable to reliably measure these arcs over many epochs to produce a more precise model.

\section{Discussion}
\label{sec:discussion} 

\subsection{Advantages of scintillation arcs}
The analysis of variations in the timescale of intensity scintillations for a binary pulsar was originally suggested and attempted by \citet{Lyne84}, but it did not provide the desired accuracy because the plasma turbulence was inhomogeneous over the orbit of the first pulsars tested. The method had more success with relativistic binaries because their compact orbits appear entirely within the region of scattering on the sky, meaning the scintillation timescale is primarily controlled by the velocity of the line of sight. This was first demonstrated by \citet{Ord+02a} with PSR~J1141$-$6545 and later by \citet{Ransom+04} with PSR~J0737$-$3039A. However both of these velocity models had neglected the anisotropy of the scattering. It was later shown by \citet{Rickett+14} that only five independent parameters can be defined by orbital variations, and the inclusion of anisotropy exceeded this limit. If the observations are extended to include the orbital period of the Earth, then this limit can be greatly extended and a complete solution is possible. Annual variations for PSRs J0737$-$3039A and J1141$-$6545 were modelled by \citet{Rickett+14} and \citet{Reardon+19} respectively to uniquely solve the three-dimensional orbits of these pulsars.

The use of the curvature of a scintillation arc, instead of the timescale, eliminates the problem of inhomogeneities in the IISM because the arc curvature is independent of the strength of scintillation, where the main effect of inhomogeneities appears. Thus including the orbital motion of the Earth can be done with precision comparable with fitting the timescale over a binary orbit. 

Scintillation arcs are particularly useful in weak scintillation where they appear sharper than in strong scintillation. In weak scintillation both the scintillation bandwidth and timescale increase and are typically comparable with the observing bandwidth and the observing time respectively. Thus there may be only a few ``scintles" in a dynamic spectrum, which makes any measurement of a correlation function difficult. However arcs originate from a finer-scale pattern caused by much higher scattering angles. Therefore, there are many more degrees of freedom in a secondary spectrum and the accuracy is correspondingly increased. 

From our observations, we see that the apparent properties of the IISM sampled by our line-of-sight (velocity, distance, and anisotropy) change slowly enough for the curvature to remain stable over many years, which gives us clean annual variations for precise transverse velocity modelling.

\subsection{Screen distances and IISM velocity}
\label{sec:distanceandvelocity}
We have measured $|V_{\rm IISM}|\gtrsim 32$\,km\,s$^{-1}$ for the primary screen and $|V_{\rm IISM}| \sim 50$\,km\,s$^{-1}$ for the secondary. We know so little about the origin for this plasma that it is difficult to say whether this velocity is unusual. It would be large compared with the thermal or Alfv\' en speed of the interstellar plasma \citep{Goldreich+95}, but an entire cloud could be moving with this velocity without causing any internal disturbance. Anything fast moving would cause shocks, which we may be preferentially seeing in the data since the density and level of turbulence increases, which then results in more scattering.

Taking these velocities into account, we were able to make robust screen distance measurements of $s=0.427 \pm 0.002$ and $s = 0.212 \pm 0.016$ for the primary and secondary arcs respectively. Using the precise measurement of the distance to PSR~J0437$-$4715 from \citet{Reardon+16}, $D=156.79 \pm 0.25$\,pc, the absolute distances to these screens are $D_1 = 89.8 \pm 0.4$\,pc and $D_2 = 124 \pm 3$\,pc respectively.

The primary screen distance is significantly different from that reported in the earlier PSR~J0437$-$4715 arc analysis of \citet{Bhat+16}. This is because they had only two observations and could not include the additional parameters required to describe any IISM velocity or scattering anisotropy and assumed a static isotropic medium. Our results show that screen distance estimates from individual arcs in single observations are unreliable. 

We also note that the position angle of the IISM velocity and the anisotropy angle estimates for the two screens are very similar (Table \ref{tab:Params}), despite being separated by $\sim 33$\,pc. This similarity is rather improbable, but with only two screen measurements we cannot determine if this is coincidence or suggestive of an association between these two screens. Future high signal-to-noise observations could permit measurements of more arcs and clarify the relative velocities of the population of screens.

\subsection{Object candidates}
\label{sec:candidates}
The structures in the IISM that cause scintillation are poorly understood because they are difficult to study. Compact, turbulent, and over-dense regions of electron density in the IISM are known to have a high scattering efficiency that can dominate the scattering of the entire line-of-sight \citep{Coles+15}, meaning that the scintillation can often be described by a single thin screen scattering model. However the origin of such compact regions, including extreme scattering events \citep[ESEs; e.g.][]{Fiedler+87, Fiedler+94, Coles+15, Bannister+16, Kerr+17}, largely remains unknown. Often plasma confinement by magnetic fields in the IISM is invoked to explain the observed scattering phenomena \citep[e.g.][]{Goldreich+95}. 

Highly anisotropic scattering has previously been suggested to originate from inclined, corrugated ``reconnection current sheets" in the IISM \citep{Pen+14}. These current sheets form at the boundaries between magnetic field configurations after they relax from an energetic disturbance such as a supernova. Other potential structures include the boundaries of local interstellar clouds \citep[e.g.][]{Linkey+08}, or the shock-heated ionized surfaces of small, self-gravitating molecular clouds \citep{Walker07}. However, the latter suggests discrete AU-scale clouds, which is a more appropriate model for ESEs \citep{Walker+17} than the sustained, stable scattering that we observe for this pulsar. 

\citet{Bhat+16} suggested that the primary screen for PSR~J0437$-$4715 may be associated with the edge of the Local Bubble (a local region of plasma under-density and high temperature), which is estimated to be at $\sim 100-120$\,pc \citep[e.g.][]{Spangler09}. Our updated distance for the primary screen is significantly less than the measurement of \citet{Bhat+16}, but either of the screens (at $89.8 \pm 0.4$\,pc and $124 \pm 3$\,pc) could be associated with the boundary of the Local Bubble.

The observed power distribution in our secondary spectra does not clearly identify discrete structures in a transverse direction, since there are no arc asymmetries, inverted arclets, discontinuities, or other clear deviations from a simple parabola that passes through the origin. The small but persistent anisotropy that we infer from the Doppler profiles and the curvature model fit to the primary arc, may simply arise from anisotropic Kolmogorov turbulence, which could indeed persist for hundreds of AU. The Delay profile in Figure \ref{fig:norm_sspec} (right panel), as well as the dispersion measure variations for this pulsar \citep{Keith+13}, are also consistent with Kolmogorov turbulence. 

\subsection{Future work}

We have identified additional faint arcs; one at a lower curvature and one at a higher curvature in a few observations. However we were unable to reliably measure the curvature for these arcs in multiple observations to find curvature modulations. This is because the arc with smaller curvature is generally faint and near to the power on the leading-edge of secondary and/or primary arc. The arc with higher curvature is hidden mostly within the power inside the primary arc.

Future observations can be optimized for fitting any arcs with higher curvature by taking long observations, and lower curvatures can be probed by using wide observing bandwidths or detecting flux in shorter sub-integration times. These additional arcs (and potentially more) will be analysed using more sensitive pulsar observations from the MeerKAT \citep{Bailes+18} radio telescope, and the ultra wide-band receiver of the Parkes radio telescope \citep{Hobbs+19}.

In further studies on the Doppler profiles it will be possible to develop templates for matched filtering to improve arc fitting measurements and to determine anisotropy and orientation simultaneously with a velocity model. This technique will become increasingly important for high signal-to-noise and wide-bandwidth observations, however it requires an estimate of the scattering strength and currently assumes that the scattering originates from a thin screen with a power spectrum of density irregularities (such as Kolmogorov turbulence). 

\section{Conclusions}

Measuring the curvature of scintillation arcs as they change with the velocity of the line-of-sight through the scattering medium is a powerful way of precisely measuring properties of the scattering and the orbit of a binary pulsar. We have measured annual and orbital modulations in the curvature of two separate arcs for the millisecond pulsar, PSR~J0437$-$4715. The two arcs correspond to distinct scattering screens, and we have precisely measured their distance and transverse velocity under two geometrical models. We find that the primary and secondary screens are located at $D_1=89.8 \pm 0.4$\,pc and $D_2=124 \pm 3$\,pc from the Earth respectively. The advantage of our long-term model is that these measurements are nearly model-independent.

We found that the kinematics of the IISM in each screen are very well modelled by just two parameters to describe the motion of the screens ($v_{\rm IISM, \alpha}$ and $v_{\rm IISM, \delta}$ for isotropic scattering, or $\xi$ and $v_{\rm IISM, \xi}$ for anisotropic) across the whole $\sim 16$\,years of observations, meaning that these properties of the interstellar plasma remain stable over a spatial scale of at least $\sim 400$\,AU. 

Our precise velocity model has provided a measurement of the longitude of the ascending node for PSR~J0437$-$4715 independently of pulsar timing, and we find $\Omega = 206.3 \pm 0.4$, which is more precise than that obtained from the timing model of \citet{Reardon+16}. Our data are also sensitive to the orbital inclination angle, and we have measured $i=137.1\pm 0.3^\circ$. This shows that modelling these variations gives an alternate means for obtaining precise measurements of $i$ for pulsars that are not observed edge-on. When applied to other pulsars that have only one post-Keplerian parameter measured from timing, this could lead to more measurements of neutron star masses, depending on the pulsar's flux and scintillation properties (from the pulsar catalogue \citep{Manchester+05}, approximately 10 pulsars are currently suitable for this application). Measurements of $i$ and $\Omega$ can also improve tests of general relativity by allowing the correction of kinematic effects in relativistic parameters \citep{Kopeikin95, Kopeikin96}, or by enabling tests of gravitational symmetries \citep[e.g.][]{Zhu+19}. Similarly to the screen distance, these binary parameters are nearly model-independent.

If the proper motion of a pulsar is not known from pulsar timing, this method could be used to estimate it by assuming that its velocity is much larger than any IISM velocity. However we have shown that care needs to be taken when estimating parameters of the screen under the assumption of a stationary IISM, since it may have a substantial (of order a few tens of km\,s$^{-1}$) magnitude. Arc curvature modelling is promising for pulsars observed in the weak scintillation regime, where the scintillation bandwidth and timescale is too unstable from observation-to-observation to reliably measure the properties of diffractive scintillation and their change with time.

\section*{Acknowledgements}

Parkes radio telescope is part of the Australia Telescope, which is funded by the Commonwealth Government for operation as a National Facility managed by CSIRO. M.B., S.O., R.M.S., and R.S.\ acknowledge Australian Research Council grant FL150100148. Parts of this research were conducted by the Australian Research Council Centre of Excellence for Gravitational Wave Discovery (OzGrav), through project number CE170100004. Work at NRL is supposed by NASA. This research has made use of NASA's Astrophysics Data System. 


\bibliography{J0437_scintillation_arcs.bib} 

\appendix

\section{Accessing data and reproducing results}

The raw pulsar observations are available for download from the CSIRO data access portal (DAP), with the majority of these data being taken under the ``P456" project code (\url{https://data.csiro.au/dap/search?q=P456}). The observations were processed using the processing pipeline for the PPTA second data release, as described in \citet{Kerr+20}. The pulsar ephemeris used to compute the pulsar's transverse velocity is from \citet{Reardon+16} and available at \url{https://doi.org/10.4225/08/561EFD72D0409}. Finally, all processed dynamic spectra of PSR~J0437$-$4715 are available for download from the DAP at \url{https://doi.org/10.25919/5f3cd2bc1c213}.

The analysis made use of a Python package we call \textsc{Scintools}, which will be described and documented in detail in a future work. The package is available from \url{https://github.com/danielreardon/scintools} and includes some example scripts that can be used to reproduce our results. These examples show the techniques for dynamic and secondary spectrum processing, arc curvature measurement, and modelling curvature measurements with time. The current version as of this publication will be preserved as ``pre-release version 0.2". The code makes use of \textsc{Astropy} \citep{Price-Whelan+18} to calculate the transverse velocity of the Earth with respect to the pulsar, and the Romer delay to the Solar System Barycenter (SSB). 

We have also reproduced the dynamic spectrum simulation code of \citet{Coles+10}, and include this in \textsc{Scintools}.

\section{Additional equations and figures}

The secondary spectrum in weak scintillation (equation D5 in \citet{Cordes+06}) can be written in terms of the spatial spectrum of the phase shift the wave experiences in traversing the scattering region $P_\phi(\vec{\kappa})$
\begin{equation}
    S(f_\lambda,f_t)=\frac{8 \pi^2}{V_{\rm eff} D_e \kappa_y} \left[ P_\phi (\kappa_x = 2 \pi f_t / V_{\rm eff}, \kappa_y) + P_\phi (\kappa_x = 2 \pi f_t / V_{\rm eff}, -\kappa_y)\right].
\end{equation}
Here the x-axis is in the direction of $\vec{V}_{\rm eff}$, $\kappa_x$ and $\kappa_y$ are spatial wavenumbers, $D_e = D s (1-s)$, $\kappa_y = \sqrt{\kappa_0^2 - \kappa_x^2}$, and $\kappa_0^2 = 8 \pi^2 |f_\lambda|/D_e$. We then define $P_\phi$ as a power-law in a quadratic form
\begin{equation}
    P_\phi(\vec{\kappa}) = C_\phi^2 / Q(\vec{\kappa})^{\alpha/2}
\end{equation}
where the exponent $\alpha = 11/3$ for a Kolmogorov spectrum.
The quadratic form Q for an axial ratio $A_r$ and orientation $\psi$ with respect to the $\vec{V}_{\rm eff}$ is \citep{Reardon+19}
\begin{equation}
    Q(\vec{\kappa}) = a\kappa_x^2+b\kappa_y^2+c\kappa_x\kappa_y 
\end{equation}
where $a= \cos^2{\psi}/A_r + A_r\sin^2{\psi}$, $b= A_r\cos^2{\psi} + \sin^2{\psi}/A_r$ and
 $c = 2\sin{\psi}\cos{\psi}(1/A_r - A_r)$.
The secondary spectrum is then separable
\begin{equation}
    S(f_\lambda,f_t) = \left( \frac{8\pi^3 C_\phi^2}{V D_e}\right) \kappa_0^{-\alpha+1} D_t(f_t/f_{arc})
\end{equation}
for $|f_t| < f_{arc}$ and 0 elsewhere. The Doppler profile is given in normalized Doppler $f_{tn} = f_t/f_{arc} = \kappa_x/\kappa_0$ so it is dimensionless
\begin{align}
\label{eqn:Doppler_prof}
    D_t(f_{tn})&=[(a f_{tn}^2 + b (1-f_{tn}^2) + c f_{tn} (1-f_{tn}^2)^{1/2})^{-\alpha/2} \nonumber \\
    &+\  (a f_{tn}^2 + b (1-f_{tn}^2) - c f_{tn} (1-f_{tn}^2)^{1/2})^{-\alpha/2}] (1-f_{tn}^2)^{-1/2}.
\end{align}
If the medium is isotropic $a = b = 1$ and $c = 0$ so $D_t(f_{tn}) = (1-f_{tn}^2)^{-1/2}$.

\begin{figure*}
\centering
\includegraphics[width=.85\textwidth]{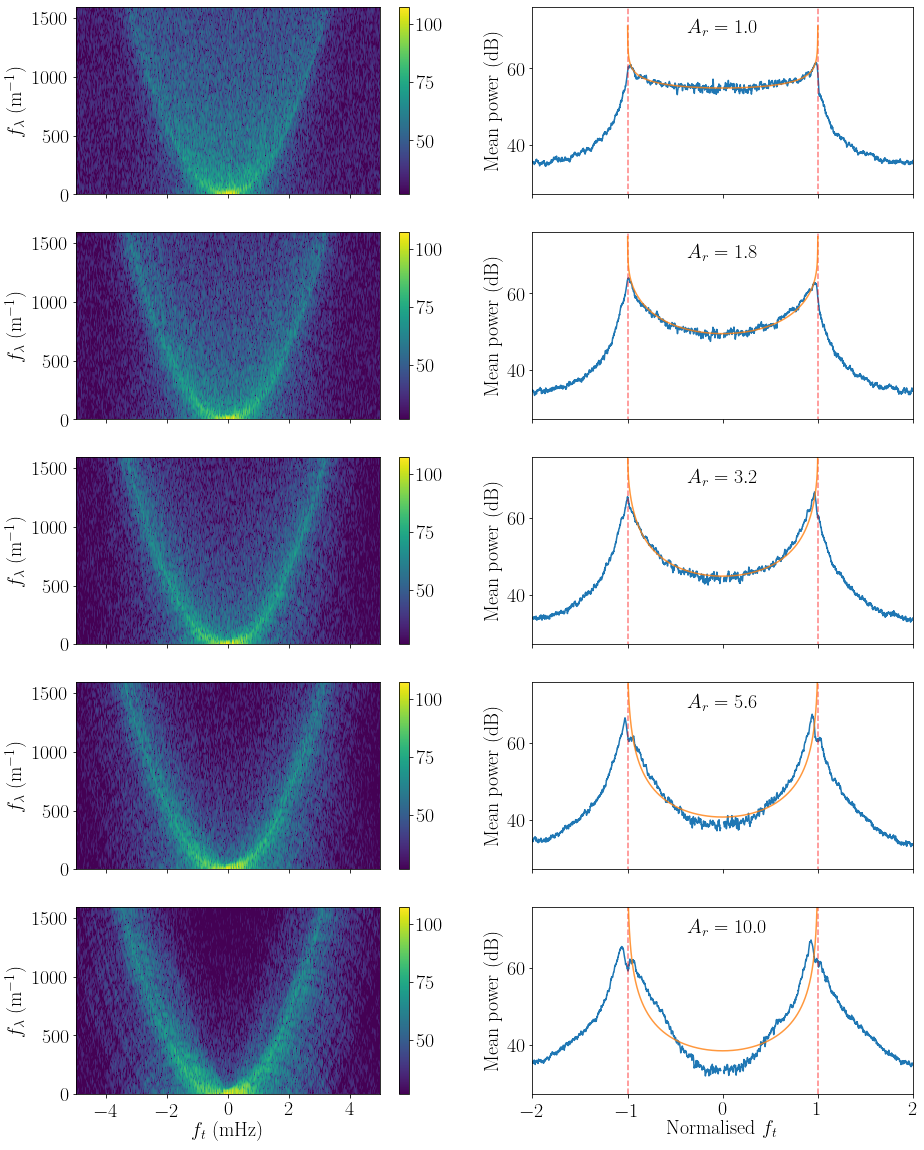}
\caption{Secondary spectra (left panels) and corresponding Doppler profiles (right panels) for a series of simulations with varying degrees of anisotropy. The axial ratios ($A_r$; labelled on the Doppler profiles in the right panels) for the simulations are spaced log-uniformly from isotropy to a 10:1 anisotropy, each one aligned with the velocity. The strength of scintillation and the sampling characteristics for the simulations were chosen to approximately match our observations of PSR~J0437$-$4715 at 20-cm. On the right panels we also show the theoretical expectation for the Doppler profiles in weak scintillation (orange lines, scaled approximately in amplitude to aid visualisation). At $A_r=10$, the appearance of inverted arclets broadens the arc, which adds power internally to the arc, which is not expected in the weak scattering regime and therefore the model is a poor approximation.}
\label{fig:anisotropy_sims}
\end{figure*}

\begin{figure*}
\centering
\includegraphics[width=.85\textwidth]{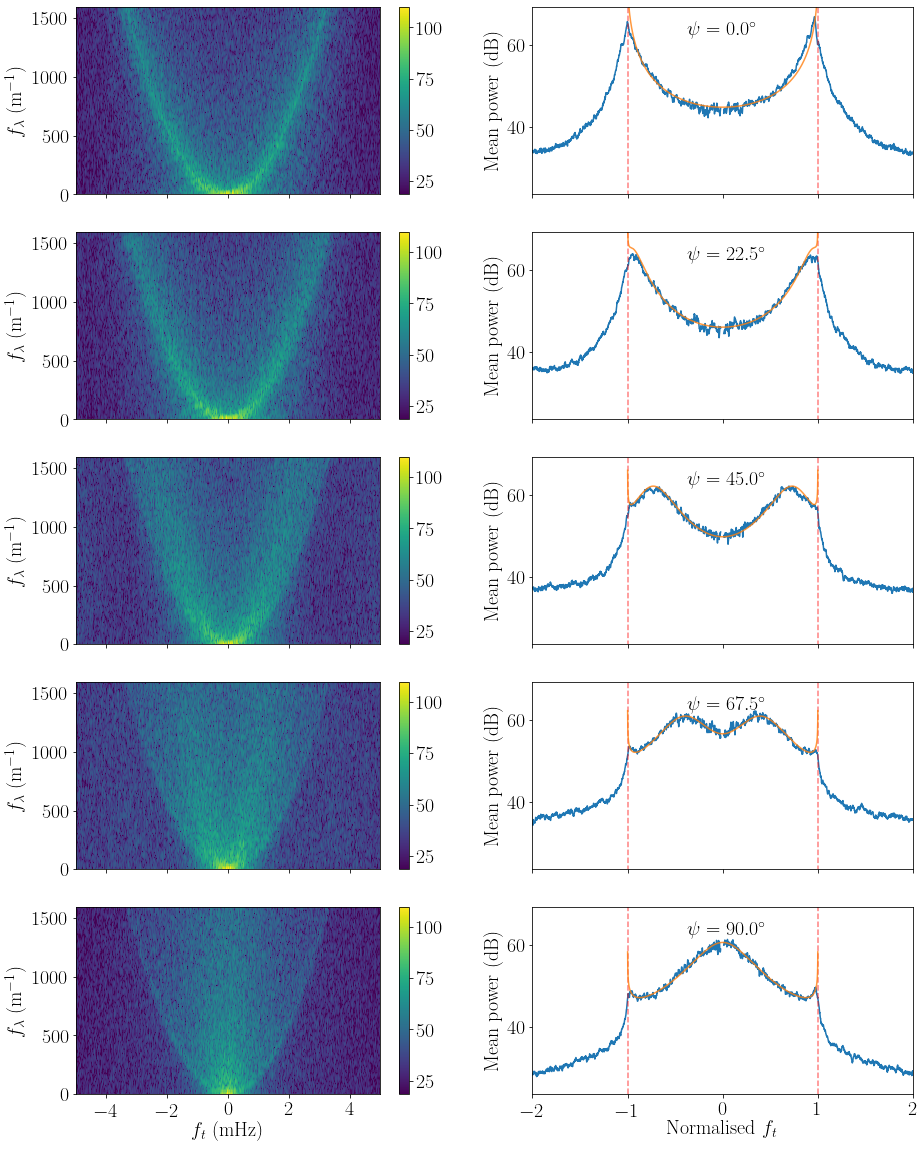}
\caption{As in \ref{fig:anisotropy_sims}, but with varying orientation of the anisotropy $\psi$ with respect to the velocity, for a fixed axial ratio of $A_r = 3.2$.}
\label{fig:orientation_sims}
\end{figure*}

We use this Equation to plot the theoretical lines in Figures \ref{fig:norm_sspec}, \ref{fig:example_obs}, \ref{fig:anisotropy_sims}, and \ref{fig:orientation_sims}. For Figures \ref{fig:norm_sspec} and \ref{fig:example_obs} we show the expectation for purely isotropic scattering to guide the eye and demonstrate that the data at these epochs is consistent with near-isotropic scattering if the majority of power inside the arc does indeed originate from the primary scattering screen itself, rather than from other scattering material along the line-of-sight. In Figures \ref{fig:anisotropy_sims}, and \ref{fig:orientation_sims} we overlay the weak scintillation approximation for our simulations of near-weak scintillation that match the properties of 20-cm observations for PSR~J0437$-$4715. This shows that for $A_r \lesssim 3$ the weak scintillation approximation holds well, and beyond this limit it fails because of the appearance of inverted arclets.

\begin{figure*}
\centering
\includegraphics[width=.6\textwidth]{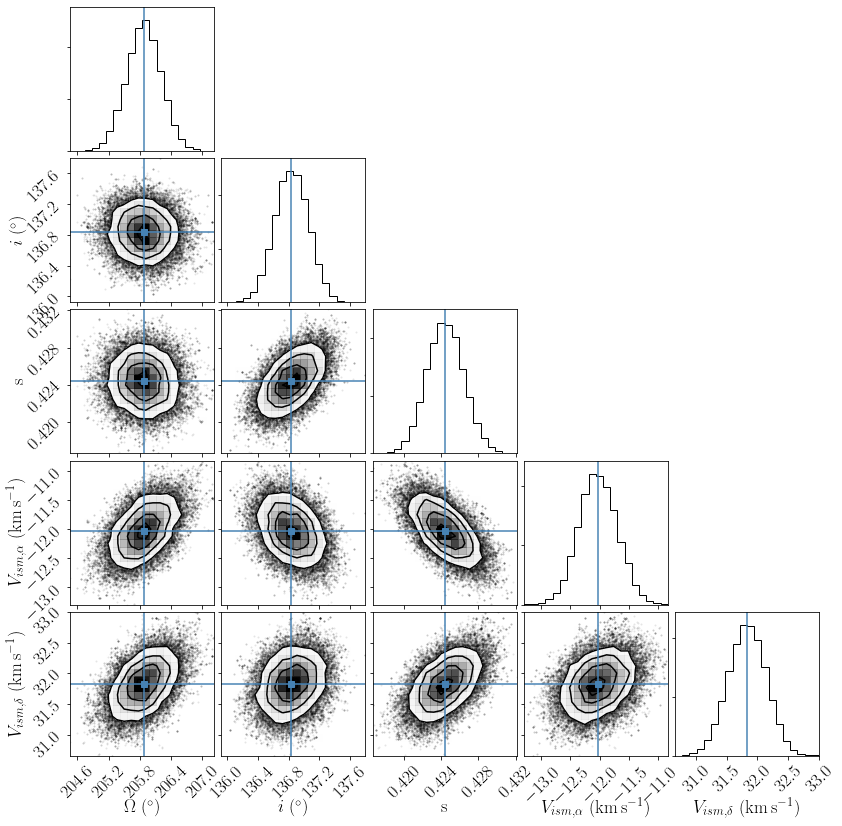}
\includegraphics[width=.6\textwidth]{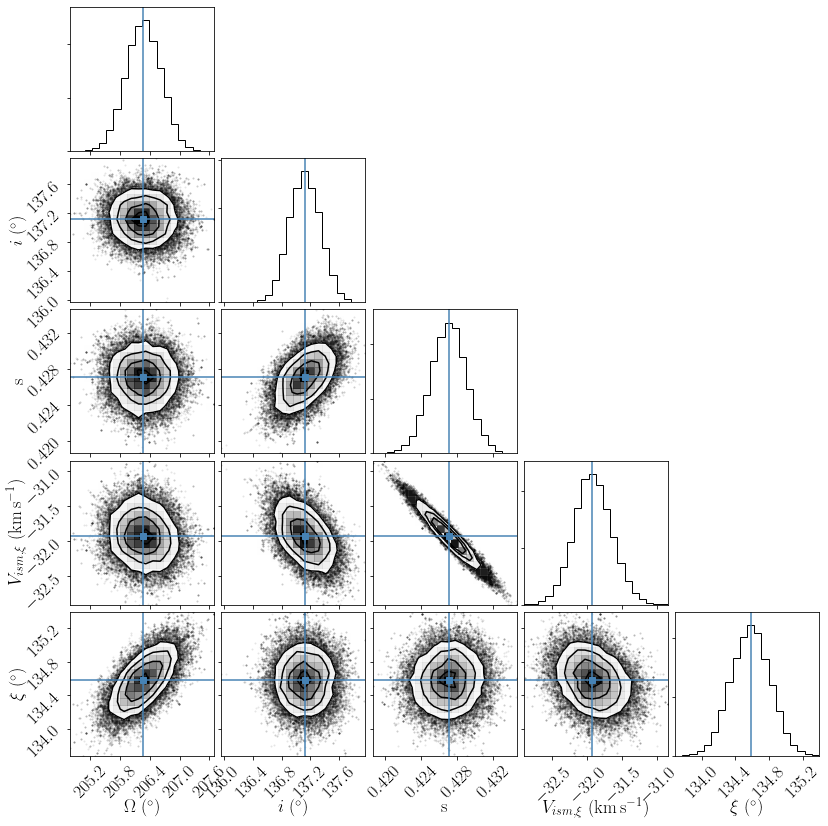}
\caption{Posterior probability distributions for the parameters in the isotropic (top) and anisotropic (bottom) models for the primary arc. The contours in the 2D distributions mark the 68\%, 95\% and 99.7\% confidence intervals and the blue lines mark the mean for each parameter. This figure was created with the \textsc{corner} package \citep{corner}. The most substantial covariances are observed between $s$ and $v_{\rm{IISM},\xi}$ for the anisotropic model with a correlation coefficient of $-0.96$, and $s$ and $v_{\rm{IISM},\alpha}$ for the isotropic model with a correlation coefficient of $-0.63$.}
\label{fig:posterior}
\end{figure*}

\end{document}